\documentclass[table,dvipsnames]{arxiv}

\usepackage{tikz}
\usepackage{xr-hyper} 

\usepackage{graphicx}
\usepackage{epstopdf}
\usepackage{booktabs}
\usepackage{siunitx}
\usepackage{enumitem}
\usepackage{verbatim}
\usepackage{dblfloatfix}
\usepackage{algorithmicx}
\usepackage{algpseudocode}
\usepackage{bbding}
\usepackage{balance}

\usepackage{amsmath}
\usepackage{amsfonts}
\usepackage{amsthm}
\usepackage{amssymb} 
\usepackage{mathtools}

\usepackage{hyperref}
\usepackage{cleveref}

\definecolor{light-gray}{gray}{0.95}

\usepackage{nimbusmononarrow}
\usepackage{orcidlink}
\usepackage{fontawesome}

\usepackage{multicol}

\usepackage{listings}
\usepackage{xcolor}
\usepackage{lineno}

\title{\LARGE Adaptive Intelligence: leveraging insights from adaptive behavior in animals to build flexible AI systems}

\author[ 1,2 \orcidlink{0000-0001-7368-4456}]{Mackenzie Weygandt Mathis \Envelope}

\affil[1]{École Polytechnique Fédérale de Lausanne (EPFL), Brain Mind Institute  Geneva, Switzerland.}
\affil[2]{Mathis Laboratory of Adaptive Intelligence.
\Envelope mackenzie.mathis@epfl.ch} 

\begin{document}
\onecolumn
\maketitle

\begin{multicols}{2}
[
\textbf{Biological intelligence is inherently adaptive — animals continually adjust their actions based on environmental feedback. However, creating adaptive artificial intelligence (AI) remains a major challenge. The next frontier is to go beyond traditional AI to develop ``adaptive intelligence,'' defined here as harnessing insights from biological intelligence to build agents that can learn online, generalize, and rapidly adapt to changes in their environment. Recent advances in neuroscience offer inspiration through studies that increasingly focus on how animals naturally learn and adapt their world models. In this Perspective, I will review the behavioral and neural foundations of adaptive biological intelligence, the parallel progress in AI, and explore brain-inspired approaches for building more adaptive algorithms.
\vspace{20pt}
}
]
\section*{Introduction}

Our world is built on a series of predictions made in our mind. That is, to us, the world is a construct, created conceptually from a series of millions of predictions harbored in the neural code. When we walk on the beach, we accumulate photons with our eyes, sound waves with our ears, we feel the wind on our face, and the sand shifting under our feet, and all of this comes together to give us the perception of a warm, sunny day with ocean waves crashing on the shore. However, our brain sits in the dark confines of our skull and tens of billions of neurons communicate through electrical activity in a bath of chemicals. As incoming sensory information is processed in the brain at different times, the brain imposes a lag in order to compile it, thrusting our perception into the past -- tens of precious milliseconds behind reality. Nevertheless, we need to act quickly, and our brains have evidently evolved solutions to overcome these delays. 
\medskip 

Moreover, we need to adapt to environments marked by unpredictability, continuous change, and infrequent repetition of decision-making scenarios. The uncertainty, variability, and complexity of these settings demand adaptive strategies, crucial for learning about local changes -- such as the shifting sand under our feet.
Dating back decades, researchers have developed new frameworks to understand (decompile) how this global prediction system may work~\cite{wolpert_internal_1995,scott_optimal_2004,todorov_optimal_2002,Friston2021WorldML}. A core tenet has been that in order to overcome sensory-processing delays we must build internal models of the world (also called world models). We use the models to make predictions about the sensory and motor consequences of our actions~\cite{Rao1999PredictiveCI,Rao2024AST,mathis_somatosensory_2017,Takei2021TransientDO}. But it remains largely unknown how the brain enables adaptive behavior, and how we can take inspiration from the brain to build more adaptive artificial intelligence (AI) is of growing interest~\cite{Sternberg2019ATO,Grossberg2020APT,Hassabis2017NeuroscienceInspiredAI}. Specifically, AI here broadly refers to systems or machines that aim to mimic biological intelligence by performing tasks that typically require cognition, such as problem solving, decision making, learning, and language processing. Technically, it encompasses machine learning (ML), deep learning, natural language processing (NLP), and computer vision. In this Perspective, I review progress in studying biological intelligence and use this to propose how we can enhance AI-based agentic systems.

\begin{figure*}[t]
    \includegraphics[width=\textwidth]{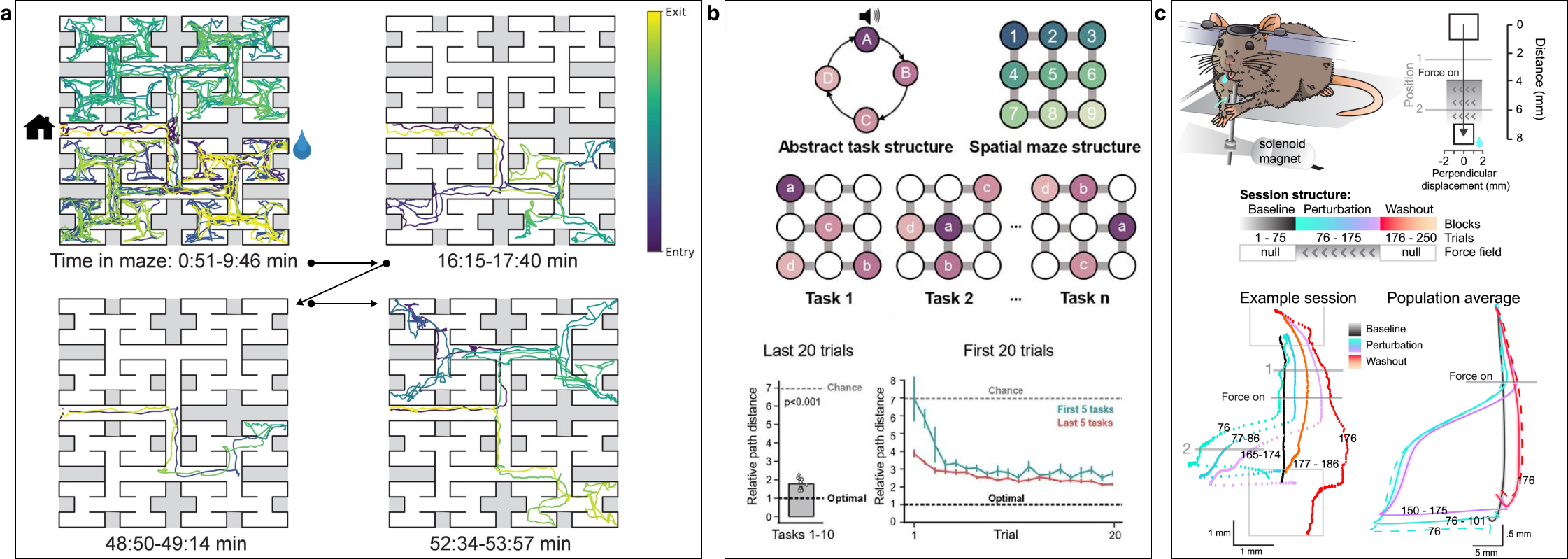}
    \vspace{1pt}
    \caption{\textbf{Rapid Learning in animals: from few-shot to updating of internal model-based learning.}
    \textbf{(a)} Adapted from~\citealt{Rosenberg2021MiceIA}: Four sample bouts from one mouse (B3) into the maze at various times during the experiment (time markings at bottom). The trajectory of the animal’s nose is shown; time is encoded by the color of the trace. The entrance from the home cage and the water port are indicated in the top left panel.
    \textbf{(b)} Adapted from~\citealt{El-Gaby2024nature}: Task design: animals learned to navigate between 4 sequential goals on a 3×3 spatial grid-maze. Reward locations changed across tasks but the abstract structure, 4 rewards arranged in an ABCD loop, remained the same. 
    Bottom left: When allowed to learn across multiple sessions, animals readily reached near-optimal performance in the last 20 trials, as demonstrated by comparing path length between goals to the shortest possible path. Bottom right: Performance improved across the initial 20 trials of each new task. This improvement was markedly more rapid for the last 5 tasks compared to the first 5 tasks.
    \textbf{(c)} Adapted from~\cite{mathis_somatosensory_2017}: Top Diagram of the joystick adaptation task structure: 75 baseline trials, 100 perturbation trials, and 75 ``washout'' (i.e., null field) trials. Bottom Left: an example session showing the average baseline trajectory (black), the first perturbation trial (sea green), average of first ten (green), and the last ten perturbation trials (purple), the first ten washout trials (orange) and the first washout trial (red). The numbers indicate trials. Right: average trajectories for all sessions (n = 27 sessions, from n = 7 mice) with temporally aligned averaging. Dashed sea green line is the average first perturbation, and the dashed red line is the average of the first five washout trials. Solid lines are average paw path ± SEM in the direction of the perpendicular deviation, with the same color scheme as the left panel; solid green is the first 25 trials during the perturbation epoch, solid purple is the last 25 trials of the perturbation epoch, and solid red line is the average of the first 25 washout trials.}
    \label{fig:animals-adapt}
\end{figure*}

\section*{Internal models for adaptive behavior}

Dating at least as far back as Aristotle, there has been a philosophical and scientific question of how our brains build models of the world~\cite{Mathis2022TheNC}. How can we combine sensory information into perception? How long does this take? Are we always tens or even hundreds of milliseconds behind reality? How could our brains make predictions about the world in order to act faster? Why do we mostly notice when the world violates our predictions? – if the ocean were suddenly quiet, we would immediately worry that we had lost our hearing and not doubt that the ocean had completely changed. While this is familiar in everyday life, we need to turn this experience into a controllable task to study it in the laboratory.
\medskip

What is the neural basis of this adaptive behavior? This historically has been difficult to answer because we are always under-sampling the neural data that underlies the behavior and often want repeated measures. The mouse brain has an estimated 70 million neurons, yet even the most state-of-the-art technology allows for recording up to 1 million individual neurons at a time~\cite{Manley2024SimultaneousCD} (and on the order of only 3Hz resolution), while most labs can record in the order of hundreds to thousands of neurons with two-photon microscopy or high-density arrays such as Neuropixels~\cite{siegle2019survey,Stevenson2011HowAI}. Thus, classically, to study how a neuron encodes a given stimulus the animal is repeatedly subjected to the stimulus and then we build encoding and decoding models~\cite{Mathis2024CELL} to gain insight into what the neural activity represents.
\medskip 

However, if you want to understand how the brain learns, there is an inherent problem with this repeated-trial approach. As Heraclitus infamously quipped, ``No man ever steps in the same river twice, for it's not the same river and he's not the same man''. Beyond the clear changes that can occur across trials during learning, the neural code can change even in steady-state given that the output behavior is rarely truly identical. Thus, averaging across trials likely misses key principles of neural computation. Since behavior is never exactly repeated anyhow, investigating learning — where behavior changes in predictable and measurable ways due to task design — may offer crucial insights into the neural basis of adaptive behavior. Namely, we can control the environmental factors that drive learning and examine the neural code across time. This not only lends itself to more ecologically-relevant tasks (the animals must be able to learn), but it's becoming a tractable approach given modern technological advances in large-scale neural recording and joint neural and behavioral analysis~\cite{Lecoq2019WideFD, Urai2022LargescaleNR,Chen2023RecentDI,Mathis2024CELL}.

\section*{Adaptive animal and neural behavior} 

There has been a concurrent push for experimental designs that incorporates more naturalistic behaviors, more complex behaviors, and tasks that enable within-session learning. Aside from new large-scale neural recording technology, a part of this is due to the advances in markerless tracking of behavior (reviewed in~\citet{mathis2020deep}), which has made digitizing movements much more tractable. What behavioral paradigms are being developed? While I cannot cover all progress in the field, I want to highlight several new approaches. Freely moving paradigms for rodents have rapidly advanced to encompass complex stimuli such as moving (threatening) novel objects~\cite{Tsutsui-Kimura202}, real-time VR worlds~\cite{Lopes2021CreatingAC}, labyrinth mazes~\cite{Rosenberg2021MiceIA}, extended multi-area home cages~\cite{Shemesh2024StudyingDA,Hao2020FullyAM}, and 3D environments such as those used to study depth estimation~\cite{Skyberg2024NaturalVB}. In several settings they have also built trial-based assays that smartly constrain these naturalistic settings in order to be able to directly probe learning across epochs.
\medskip

An elegant thread of work across multiple groups aims to build tasks for zero- or few-shot learning in complex spatial worlds. Zero-shot learning is a term adapted from ML that describes the ability of a model to generalize to new tasks or concepts without any specific training examples~\cite{Palatucci2009ZeroshotLW}. Few-shot means that it can learn from only a few examples. For example,~\citet{Rosenberg2021MiceIA} developed a task where mice must seek rewards in a labyrinth (Figure~\ref{fig:animals-adapt}a). They found that mice can not only make up to 2000 decisions per hour, but after unsupervised exploration of the maze they could few-shot learn to find the water spout (in around 10 tries)~\cite{Rosenberg2021MiceIA}. Notably, this is a learning rate that is 1000-fold higher than two-alternative forced choice (2AFC) tasks that are typically deployed in rodent studies of decision-making, meaning this task allows one to better study rapid learning. Another example is showing that mice can zero-shot learn new sequences of rewards (such as go to location A, B, D, then C)~\cite{El-Gaby2024nature}. Mice were trained on multiple tasks with a shared underlying structure that organized a sequence of goals, though the specific goal locations varied. Strikingly, the mice learned this common structure, allowing them to make zero-shot inferences on the first trial of new tasks (Figure~\ref{fig:animals-adapt}b). Intriguingly, they also found neurons in medial prefrontal cortex that acted as task-structured memory buffers, namely, they tracked the progression along behaviorally-relevant steps~\cite{El-Gaby2024nature}. Such schemata could perhaps be mapped to memory-replay in neural networks (see below).
\medskip

Animal intelligence is, of course, not limited to laboratory animals. Dating back to early comparative cognition studies~\cite{thorndike1898} there has been a large body of work studying intelligence in a myriad of animals. Crows, for example, exhibit tool use and causal reasoning, with New Caledonian crows even manufacturing tools to retrieve food~\cite{Hunt1996ManufactureAU,RUTZ2012153}. Bees demonstrate numeracy and complex communication through the waggle dance, encoding spatial information about food sources~\cite{dyer1991} and can solve ``puzzles" such as pulling a string to move a food source closer~\cite{alem2016}. Various domesticated animals also exhibit human-aligned intelligence. Horses show social learning and can interpret human emotional cues~\cite{Proops2018}. Dogs display theory of mind and word comprehension that rival that of young children~\cite{Kaminski2004dogs}. Such findings even sparked a reoccurring virtual animal AI Olympics challenge~\cite{beyret2019animalai}. 
\medskip

In sensorimotor control there has been a long line of work on rapid learning, which is called motor adaptation. For clarity, here I define learning as the acquisition of a skill (and thus learning a new internal model), while adaptation is using a learned internal model and updating it. Motor adaptation studies have been developed where visual or proprioceptive information is perturbed such that within-session, an animal must adapt. This has not only spurred the development of new neural analysis tools~\cite{Williams2017UnsupervisedDO,Sorscher2022NeuralRG,Schneider2023LearnableLE, Mathis2024CELL}, but a host of behavioral tasks. For example, visuomotor rotations have been used in humans and non-human primate studies to study how they can account for sensorimotor discrepancies in only a few hundred trials~\cite{mcdougle_explicit_2015,krakauer_human_2011,izawa_learning_2011,Williams2017UnsupervisedDO,Stavisky2017TrialbyTrialMC}. The same principle is used for changing environmental dynamics of a manipulandum that causes deviations in a limb-movement tasks~\cite{shadmehr_adaptive_1994,bizzi_arm_2013,mathis_somatosensory_2017, DeWolf2024}. Notably, these tasks are specifically designed to measure both adaptive learning, and the formation of internal models. There is always a period of baseline control movements, perturbations, and a return to baseline-condition that allows researchers to behaviorally measure if an internal model was updated (Figure~\ref{fig:animals-adapt}c). 
\medskip

Neurally, evidence shows that motor and sensory areas can change their tuning properties during the course of learning~\cite{li_neuronal_2001,Sun2022CorticalPA,DeWolf2024,Meyer2018SingleexposureVM,Meirhaeghe2021APA}. A study by~\citet{Sun2022CorticalPA} showed that during a motor adaptation (learning) task in macaques, motor cortex (M1) can ``index'' memories of the hand-force required for the process of learning to adapt in the form of movement readiness potentials that occur prior to movement execution. Notably, the neural subspace most predictive of hand forces changed during the period before movement (preparatory), specifically during the learning epoch. In a neural dimension orthogonal to this force-predictive subspace, they identified a uniform shift across all movement directions, including those unaffected by learning. Intriguingly this uniform shift remained after exposure to the force field, reflecting an updated internal model.
In addition, other works also show new evidence of state-changes that persist across learning~\cite{DeWolf2024,Takei2021TransientDO}, and temporally-resolved prediction errors~\cite{keller_sensorimotor_2012,DeWolf2024}. 
Influential work on visual cortex (V1) has shown the existence of `mismatch' neurons when the expected visual feedback is disrupted~\cite{keller_sensorimotor_2012}, and similar prediction errors have been found in somatosensory (S1), motor (M1), and frontal areas of cortex~\cite{Meirhaeghe2021APA,DeWolf2024}.
\medskip

Some of the best evidence for causally showing how neurons adapt has come from brain-machine-interface (BMI) studies that require the subject to directly alter neural firing in order to change something in the external world (like a cursor on a screen).
Closed-loop systems, like calcium-based BMIs (caBMIs, BMIs driven by decoded optical activity readout of calcium fluorescence signals in M1)  or with electrical stimulation, have been used to study the timescale and subsets of neurons that can be used, as not all neuronal types have been found to be equally adaptable. Fundamental work has revealed that even small numbers of neurons can be leveraged for decoding, and for BMI-guided feedback~\cite{Williams2017UnsupervisedDO,Sadtler2014NeuralCO}. Notably, they often live in discrete subspaces of the neural dynamics. \citet{VendrellLlopis2021DiverseOC} pushed this further to link these subspaces to cell-types. They trained mice to modulate the activity of either intratelencephalic (IT) or pyramidal tract (PT) neurons for reward. They found that mice learned to control PT neuron activity more quickly and effectively than IT neuron activity. This intriguingly could be related to the anatomical connectivity and differing inputs.

\begin{figure*}[t]
\centering
    \includegraphics[width=\textwidth]{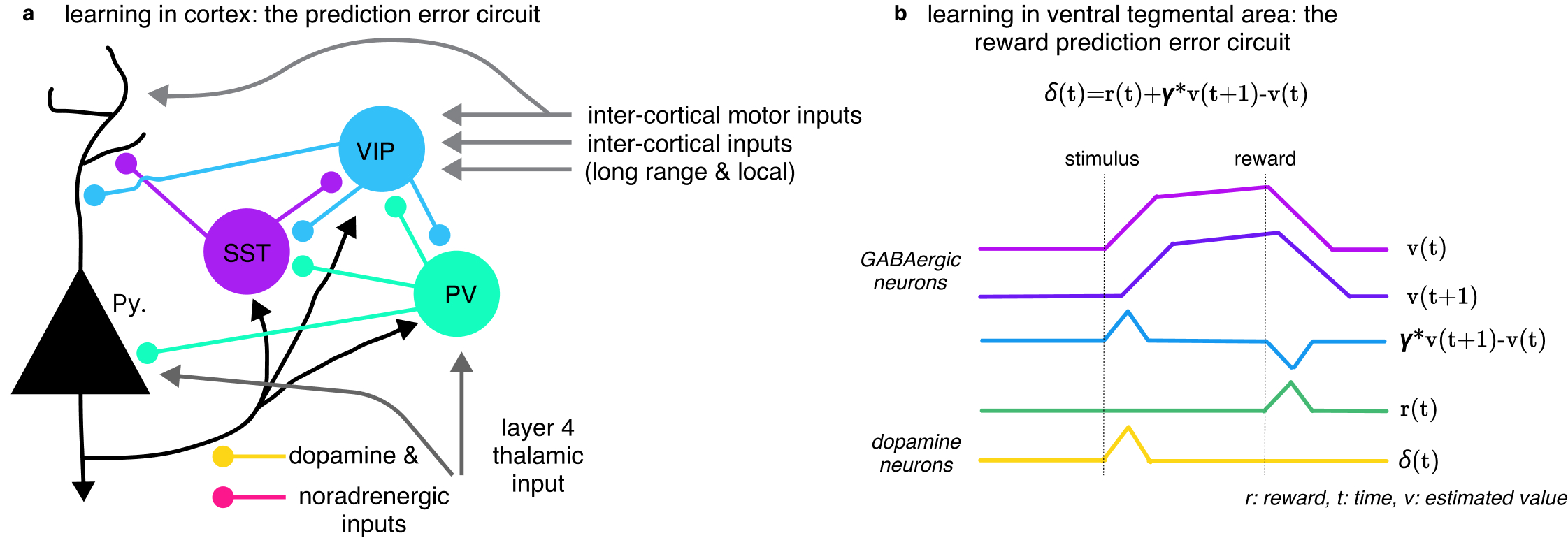}
    \caption{\textbf{Neural computations: biological teaching signals.}
    \textbf{(a)} Diagram of the theoretical prediction error circuit (based on evidence from~\citet{keller_sensorimotor_2012,LEINWEBER2017,Green2023ACE,Jordan2023TheLC}). 
    \textbf{(b)} Diagram of the computation in the reward prediction error circuit of the midbrain (ventral tegmental area) dopamine and GABAerigc neurons. Inspired from~\citet{Gershman2024ExplainingDT,cohen_neuron-type-specific_2012}. }
    \label{fig:circuits}
\end{figure*}

\section*{Anatomical links to computations}

The anatomy of the brain is important to consider, as this too could have direct implications for the future design of adaptive AI systems. Several areas in the brain have this incredible structure, -- layers in the cerebral (neo-) and cerebellar cortex -- which could be a critical part of new architectures for AI systems. There is a long-standing debate about representation emergence from data vs. architecture constraints~\cite{Lake_Ullman_Tenenbaum_Gershman_2017,Raghu2021}, but it is increasingly clear that architecture is critically important and linked to function. Fundamental work dating back to the 1970’s is worth revisiting. Vernon Mountcastle describes a framework of the organizational principle of the neocortex~\cite{Mountcastle1978}. He delineates evidence that the major functional division in the neocortex is not whether an area is ``sensory'' or ``motor,'' but rather the vertical neocortical column constitutes the basic computational unit, and the input-output pattern merely dictates the space of information it acts upon—namely, the ``auditory'', ``visual'' or ``motor neocortex'' has the same cellular scaffold and should only considered a particular region based on the type of sensory input. Nonetheless, this architectural bias gives rise to computations on information with brain area-specific information.
\medskip

Form ties to function, and one emerging hot topic is how cortical circuits implement learning from prediction errors across cell types and layers. Building on the initial discovery of sensory prediction errors in cortex~\cite{keller_sensorimotor_2012}, subsequent works have begun to record from subtypes of neurons such as parvalbumin (PV), vasoactive intestinal peptide (VIP) and somatostatin (SST) classes of interneurons, and excitatory neurons in order to form a more complete model of how both excitatory and inhibitory neurons across layers could implement the learning rule~\cite{Green2023ACE,Wilmes_2024,LEINWEBER2017} (Figure~\ref{fig:circuits}a). This work in cortex follows earlier work that discovered reward prediction errors (RPEs) in the midbrain (within the ventral tegmental area)~\cite{schultz_neural_1997,cohen_neuron-type-specific_2012,Gershman2024ExplainingDT}, and found that they are also part of an inhibitory-gated circuit where GABAergic neurons play a crucial role in the computation of RPEs~\cite{Eshel2015ArithmeticAL} (Figure~\ref{fig:circuits}b).
\medskip 

Recent work aims to unify how a hierarchical implementation of prediction errors (including RPEs) could aid in learning across sensory, cognitive, and motor systems. \citet{Tsai2024} found that during a cue-guided reward task, layer 2/3 somatosensory neurons showed enhanced responses to reward-predictive stimuli, and this led to reduced reward-prediction errors and increased confidence in predictions. Following rule reversal, the lateral orbitofrontal cortex, via VIP interneurons, signaled context-prediction errors, reflecting a loss of confidence. The work suggests a hierarchical interaction of prediction errors across cortical regions, with top-down signals modulating sensory cortex activity
Notably, prior work finds limited roles for reward-only driven  motor learning, but it can be combined with sensory prediction errors to shape performance~\cite{izawa_learning_2011, mathis_somatosensory_2017,Palidis2021NullEO}. I envision that future work will more deeply test how various learning signals are used concurrently across systems. For example, tasks that have different types of prediction errors can be developed that might conflict, forcing the subject to weight different errors in order to guide future actions.

\section*{Neural development and inductive biases}

In the context of learning and the links to AI, another critical angle is neural development. There are natural connections to make that could also inspire new inductive biases in the form of architectures or optimization algorithms in AI systems. For example, during development cells in the cortex migrate to their final resting location at different times~\cite{Lim2018DevelopmentAF}. This process of spatial and temporal organization, I believe, has parallels in AI, particularly in how artificial neural networks undergo initialization and optimization processes to organize their nodes across layers in deep learning models. Moreover, synapses, such as those at the neuro-muscular junction of alpha motor neurons and muscles, are formed after activity-dependent pruning~\cite{Wu2010ToBA}. This mirrors the pruning techniques (so called knowledge distillation in ML~\cite{Gou2020KnowledgeDA}) used to improve model performance, where unnecessary connections are discarded during training to enhance efficiency.
\medskip

There is also a complex and highly regulated developmental program in the spinal cord involving transcription factors and HOX genes~\cite{Dasen2009HoxNA}. This genetic regulation lays out global (rostral-caudal) and even local (dorsal-ventral) connections. Similarly, the initial structure of a neural network is determined by a set of weights and biases, which are refined through training to create the connections necessary for learning. Additionally, the so-called genomic bottleneck plays a key role in determining early innate abilities of animals, from the immediate suckling reflex to the more time-intensive but impressive ability of a horse to stand and walk a few hours after birth~\cite{Shuvaev2024EncodingIA}. In AI, this is akin to pre-trained models or transfer learning, where networks leverage previously learned features to rapidly adapt to new tasks with minimal input.
\medskip

Moreover, in the cortex, early traveling spontaneous waves shape the basis of early connections between cortical neurons. These waves are thought to synchronize neural activity across different regions of the brain. Intriguingly, new work has highlighted how the transformer architecture naturally gives rise to waves, which are a critical feature in the brain~\cite{Muller2024TransformersAC}. This is reminiscent of how attention mechanisms in transformers enable dynamic, wave-like information flow across different parts of the model, helping the system maintain coherence across large datasets. Waves are seen not only in early development but also in steady-state, where they are thought to maintain global connectivity in the brain, encoding states like arousal and attention. This functional organization in the brain has direct analogs in how AI systems, like attention networks, maintain global contextual understanding across their layers.

\section*{Measuring changes in the neural code}

 New tools are critical for the study of adaptive behavior. Computational neuroscience has made large advances in the past decade, fueled by the need to better understand ensembles of neurons~\cite{Urai2022LargescaleNR,Gallego2017NeuralMF, Perich2017ANP,Pandarinath2017InferringSN} and by leveraging the concurrent innovations in deep learning, which is reviewed in~\citet{Hurwitz2021BuildingPM} and \citet{Mathis2024CELL}. In brief, algorithms such as variational autoencoders, transformers, contrastive learning-optimized neural networks, and diffusion models are already making big strides in our ability to combine data across subjects, model multi-region interactions, and `decode the brain'~\cite{Mathis2024CELL}. I want to briefly highlight several tools that enable time-series analysis in single neurons and neural populations, as these are critical for adaptive learning studies, and I suspect will become more popular tools for studying AI systems. 
 \medskip

One of the most impactful tools in the last 20 years has been the adoption of generalized linear models (GLMs) and generalized additive models (GAMs) for neuroscientific applications~\cite{Paninski2004MaximumLE,Pillow2008SpatiotemporalCA,Balzani2020EfficientEO}. Here, a model is trained using a feature matrix (can be one or many features) with the task of predicting the spikes of neurons given the set of features. 
While GLMs can come in many flavors, as an example, generally to handle variability in behavior models are trained and tested on different splits of the data (also for cross validation). Predictions from these test sets are combined, and a pseudo-R$^2$ is calculated. 
This does require that behavior is repeated, and also allows for flexibility; i.e., there is no direct trial-averaging or behavior-triggered averaging of the neural code.
\medskip

For ensembles of neurons, several methods have been used or developed, ranging from PCA for measuring neural trajectories~\cite{Jazayeri2017NavigatingTN, churchland2012neural}, contrastive learning paired with consistency metrics to measure changes in the latent neural dynamics across learning~\cite{DeWolf2024,Schneider2023LearnableLE}, to nonlinear dynamical systems for task- and task-irrelevant metrics~\cite{Sani2024DissociativeAP,Sani2020ModelingBR}. These are more deeply covered in reviews elsewhere~\cite{Hurwitz2021BuildingPM, Mathis2024CELL}, but I will note that several approaches can be used for single-trial dynamics and have been especially impactful in motor learning and adaptation studies in non-human primates~\cite{Pandarinath2017InferringSN,Keshtkaran2021ALN}. Work by~\citet{azabou2023a} introduce a training framework and architecture designed to model population dynamics in large-scale neural recordings, which tokenizes individual spikes to capture fine temporal structures and constructs a latent representation of neural population activity. They trained this model on nearly 100 hours of data and show excellent cross-task generalizing in primate reaching studies. One limitation of this approach is the requirement of labels at test-time, but in practice this is nearly always available.
\medskip 

I predict we are just on the cusp of an influx of new advances in (neuro)science due to advances in AI. For example, large-language models (LLMs) are already starting to appear in behavioral neuroscience for digitizing and quantifying  actions in video-based data~\cite{ye2023amadeusGPT}. New vision-language, hierarchical, and multi-task foundation models are surely coming~\cite{Castro2025}. For neural analysis, already several groups are actively working on building more unified encoders that can not only be used for brain-machine interfaces, but they themselves could serve as models of the brain~\cite{azabou2023a,Schneider2023LearnableLE,Zhang2024TowardsA,Benchetrit2023BrainDT,Mathis2024CELL,wang2025}. This is sure to give rise to new approaches to extract meaningful computational principles from neural dynamics, and these new learning rules have great potential to directly influence how we train, optimize, and deploy machine learning systems.

\section*{Training and learning in artificial systems}

For understanding how neuroscience is poised to influence modern AI , here I provide some background on modern AI approaches.
Today, most AI applications in production revolve around a train-test-deploy cycle, and therefore are inherently not adaptive. A dataset is curated, which can be lab-project scale (or in the case of generative pretrained transformers (GPTs), this would include nearly all the open-source content of the Internet), and then split into a train-and-test fraction. Most efforts benefit from leaving out-of-distribution (OOD) data in the test set for a realistic measure of how well the trained model generalizes. OOD is defined as data that is significantly different than the training data. If generalization is not needed -- perhaps you are training a specific model for gait analysis in a specific laboratory setting -- then the train/test split is often a random subset of the original dataset. But for many applications, training once and deploying a robust model is ideal. 
\medskip

How do we build a robust, generalizable model that can handle changes in the real-world? While the remarkable progress of GPTs in LLMs and the discovery of scaling laws has massively accelerated progress in AI, we aim for something better -- more adaptive. What has been tried? This is where, at minimum, the sub-fields of continual learning (lifelong learning) and in-context learning come into view.
\medskip

\section*{Current approaches to adapting models}

Continual learning is the task of having a neural network learn a new series of tasks of the same modality, such as computer vision tasks, over time.  It aims to address the challenge of developing models that can learn incrementally from a continuous stream of data without forgetting previously acquired knowledge. This is something that biological brains excel at - we typically don't forget older information as we learn new skills. Yet, traditional machine learning models often face catastrophic forgetting~\cite{McCloskey1989CatastrophicII} when retrained on new data, but new advances in continual learning have introduced techniques such as local module composition~\cite{NEURIPS2021_fe5e7cb6}, knowledge distillation~\cite{NEURIPS2019_83da7c53}, elastic weight consolidation (EWC)~\cite{Kirkpatrick2016OvercomingCF,Ovsianas2022ElasticWC}, synaptic intelligence, and memory-replay~\cite{Wang2022MemoryRW,Ye2024SuperAnimal} to mitigate these issues. For a more extensive discussions I point the readers to~\citealp{Wang2023IncorporatingNA,10444954}.
\medskip

Although not directly brain-inspired, Elastic Weight Consolidation (EWC) works by constraining certain parameters to reduce interference between tasks, effectively mitigating catastrophic forgetting~\cite{Kirkpatrick2016OvercomingCF}. More recent work has adapted EWC to fine-tune self-supervised models, which has yielded performance improvements on tasks that have biased datasets by maintaining knowledge of previous tasks better than traditional methods~\cite{Ovsianas2022ElasticWC,v.2018variational}. If we attempt to map this to neural dynamics, this might be related to the specialization of circuits. Namely, the ``parameters'' of certain areas could be fixed (such as primary receptors), whereas others could be more plastic (such as hippocampus and neocortex).
\medskip

On the other hand, synaptic intelligence is a learning rule deeply inspired by brain plasticity~\cite{Zenke2017ContinualLT}. The authors developed intelligent synapses that adaptively store information, minimizing the forgetting of previously learned tasks while acquiring new ones. This approach mimics biological neural networks that balance plasticity and stability, enabling continual learning in artificial neural networks. Notably, this method can perform as well as EWC but can be performed online. Now, with modern scalable computing and architectures, this type of learning rule could be pivotal for building more adaptive systems.
\medskip

\begin{figure*}[ht]
    \includegraphics[width=\textwidth]{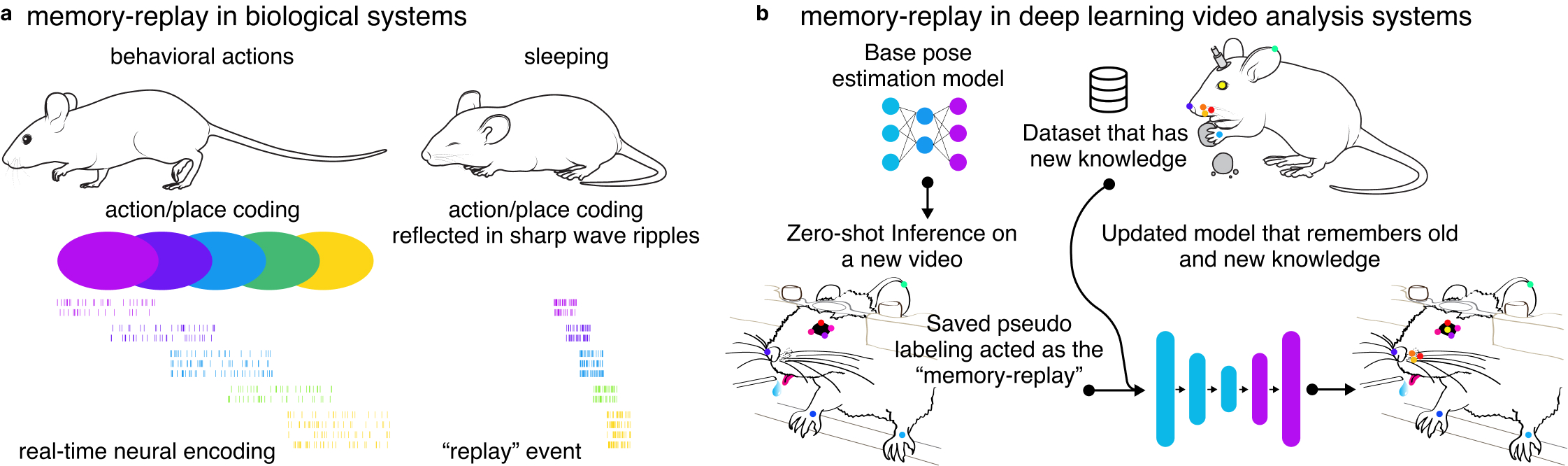}
    \caption{\textbf{Memory-replay in biological and artificial systems.}
    \textbf{(a)} In biological systems (such as in mice) neurons have been shown to encode place in their firing rates. These neurons can be active in the same sequence in sleep, which is thought to reflect memory-replay.
    \textbf{(b)} In deep learning systems one can build a replay buffer in continual learning tasks. For example, in video analysis for animal pose estimation one can use inferred labels with high confidence as pseudo-labels in order to retain base-model knowledge, even when (1) the base model's data might not be available to fine-tune, or  (2) when a new dataset, perhaps with new labels, as depicted, needs to be learned without forgetting other labels. Mouse images adapted from scidraw.io.}
    \label{fig:memory}
\end{figure*}

Another brain-inspired machine learning algorithm for learning has been the development of memory-replay~\cite{Roscow2021LearningOM,Lin1992SelfimprovingRA}, which mimics the memory systems in the hippocampus (Figure~\ref{fig:memory}a). Notably, memory-replay also, perhaps indirectly, is built on the memory-inspired system of the Hopfield Network, which is a type recurrent neural network designed for content-addressable memory, resembling a spin glass system~\cite{Hopfield1982NeuralNA}, whose invention also won John Hopfield a Nobel Prize in 2024.
Memory-replay aims to have a memory bank of already learned actions (or images or tokens, etc.) that can be ``replayed'' (interweaved) into the training data in order to limit catastrophic forgetting (Figure~\ref{fig:memory}b). This became a popular technique in reinforcement learning, and more recently, in computer vision and in LLMs. This training approach can also be used during active learning when new data (unlabeled, pseudo-labeled or labeled) can be used to fine-tune a model. If the input data continues to morph out-of-distribution (OOD) then using memory-replay can be leveraged to be sure the prior performance can be achieved. 
\medskip 

Several memory-replay-based approaches in LLMs emerged after the release of ChatGPT in November 2022 (with it's impressive language abilities yet limited content window, i.e., lack of memory), such as AmadeusGPT~\cite{ye2023amadeusGPT}, MemGPT~\cite{packer2023memgpt}, and Voyager~\cite{wang2023voyager}. AmadeusGPT introduced a short- and long-term memory where specific keywords could be quickly recalled in order to overcome token limits (i.e., the 4096 in GPT-3.5), while MemGPT used a vector database to construct a persistent memory to maintain context across interactions, and Voyager iteratively refined its skills by replaying feedback from past actions in games such as Minecraft.

\section*{Learning with spiking neural networks}

Lastly, one of the most biologically inspired and grounded advances has been in the development of spiking neural networks (SNNs) in the 1990's. Called the ``third generation of neural networks''~\cite{Maass1996NetworksOS}, SNNs are unique in their ability to model a time-dependent spike process, making them not only biologically more plausible, but also highly energy efficient and can be used on edge-computing devices. Notably, SNNs naturally lend themselves to hardware accelerated devices and neuromorphic computing.
Platforms such as Intel's Loihi, IBM's TrueNorth, and SpiNNaker provide hardware specifically designed to run SNNs efficiently~\cite{Roy2019TowardsSM}. Moreover, several early work highlights their modularity and utility for multi-area modeling~\cite{eliasmith2012large}, and how multi-area SNNs could be used with semantic pointers for multi-area weighting of information~\cite{Blouw2016ConceptsAS}.
However, historically they have been difficult to train and scale, but that is rapidly changing and new methods to transfer information from ANNs to SNNs are emerging ~\cite{He2024AnEK} and/or to directly compute the required gradients from SNNs~\cite{Wunderlich2021EventbasedBC}.

\section*{How to build more adaptive AI}

How can we take inspiration from adaptive behavior and the brain to build new AI approaches~\cite{Hassabis2017NeuroscienceInspiredAI}? A core component of adaptive behavior is having already learned priors -- these aforementioned internal models -- that can be dynamically called upon. These priors clearly require data-at-scale to form -- this ``data'' can be baked into neural circuits via the genome, or learned throughout life. An example of genetic priors would be the innate ability to suckle at birth across many species, or how a newborn horse stands and walks in a few hours after birth, while a non-genetic learned prior would be a motor skill such as playing a musical instrument. Another data source is the massive amount of unsupervised sensory and motor stimuli we rapidly accumulate across development and life. But how do we turn data into internal models?
\medskip 

While many questions remain about how these models are implemented in neural dynamics, there is increasing evidence of core computations such as the previously discussed prediction errors~\cite{Keller2024PredictivePA,keller_sensorimotor_2012} across brain regions, which shape and update internal models. Can we use these prediction errors as teaching signals in neural networks? Moreover, these prediction errors are anatomically defined, imposing architectural constraints on these systems that stem from brain-body dynamics. Can we leverage this feature in adaptive AI systems? Importantly, these prediction errors are canonical and act in specific sensory reference frameworks (such as vision or audition). As Mountcastle noted, it is less about ``sensory'' or ``motor" spaces but rather a function of the data input/output, and by extension the shared architecture and computations. Thus, as I will argue below, we should take inspiration from building specialized nodes that have core computations in the right reference framework that use prediction errors to smartly update modules, much like the brain uses prediction error to update internal models.

\section*{Foundation Models}

\begin{figure*}[hb]
    \centering \includegraphics[width=\textwidth]{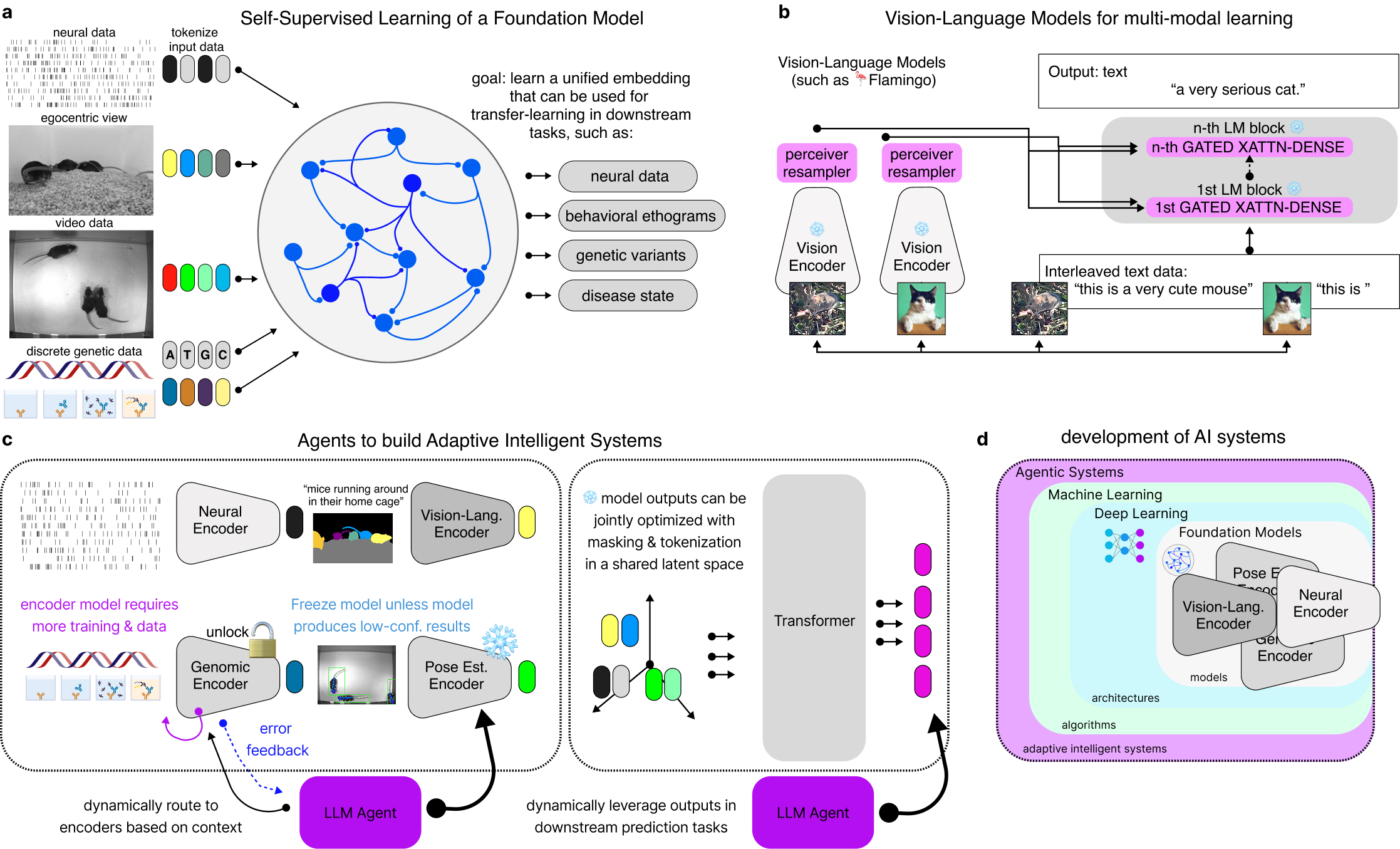}
    \caption{\textbf{Foundation Models and Adaptive Agents.}
    \textbf{(a)} One path to building generalizable models is to input diverse data, tokenize it, and train collectively with self-supervised learning. 
    \textbf{(b)} Example of Vision-Language Models (VLMs) that interweave text and images. Inspired by, and the Cat image is from,~\citet{Alayrac2022FlamingoAV}.
    \textbf{(c)} My proposal is to leverage robust generalist foundation-model encoders that a domain specific information who's outputs can be tokenized. For example, ``pose estimation" can output skeletons of animals, ``genomic" can be the DNA (or protein) sequence, ``neural" could be tokenized spikes, and ``vision-language" can be video captioning and actions (i..e, a higher level abstraction of behavior). In this example the assumption is there is a link between the genome, the animal phenotype, and the neural code. Critically though to my proposal is to then have an LLM-based agentic system oversee accuracy per encoder and pass forward only reliable outputs for joint latent-space optimization and joint decoder. In this way, the larger model need not be taken offline but rather be systematically updated when errors are detected in various encoders outputs. 
    \textbf{(d)} Building on~\citet{Bommasani2021foundation}, agentic systems can leverage Foundation Models and advances in ML and deep learning in order to move beyond the emergence of functionality and into adaptive intelligent systems.
    Genetic and neural icons adapted from scidraw.io; mouse images are from the Allen Institute and~\citet{Lauer2022MultianimalPE}.}
    \label{fig:agents}
\end{figure*}

First, let us consider what is currently the top approach for building generalizable models whose aim is often to ingest several data modalities (such as text and video) and output either. The trend in machine learning is to build unified ``Foundation Models'' that generalize to unseen data or can be used for transfer learning in downstream tasks~\cite{Bommasani2021foundation}. The Foundation Model idea for natural language processing takes large-scale unlabeled data and learns a joint representation within a model (a transformer) using tokenization and self-supervised learning~\cite{Bommasani2021foundation} (Figure~\ref{fig:agents}a). The most infamous of this type is ChatGPT and other GPT models (such as the Gemini model series~\cite{geminiteam2024geminifamilyhighlycapable} or DeepSeek~\cite{deepseekai2025deepseekv3technicalreport}). In neuroscience, several Foundation Models are also emerging for specific domains, such as for animal behavior analysis~\cite{Ye2024SuperAnimal} and for modeling the visual systems of mice~\cite{wang2025}.
\medskip 

Other emerging and prominent approaches are generalizing this approach to multi-modal  data~\cite{Alayrac2022FlamingoAV,li2024llava}. One such example is Flamingo, a Visual Language Model (VLM) that integrates pretrained vision and language models, processing mixed visual-text data and handling images or videos as inputs~\cite{Alayrac2022FlamingoAV} (Figure~\ref{fig:agents}b). This could be extended to more modalities, as recently done for robotics~\cite{Wang2024ScalingPL}. This is a rapidly advancing area of research that will surely influence neuroscience in the coming years.
\medskip

While I believe this is an exciting path for models that show better generalization, I believe we can push further to build smartly adaptive, agentic systems. We should take inspiration from our brain's specialization and build series of smartly interconnected specialist models that can then be adapted. Namely, just as all sensory/motor information gets converted to spikes -- a shared computing space this is an efficient digitization of analog signals -- notably, this information is computed upon within specific brain regions. Just as a cortical column in the primary visual cortex processes a particular region of the visual field and has specific coding properties related to visual features (like texture or edges), and the cortical circuit in the sensorimotor region computes in egocentric 3D kinematic, force, or even muscle space~\cite{Sun2022CorticalPA, DeWolf2024}.
\medskip 

Moreover, across the animal kingdom we see specialization in brains everywhere: from the remarkably high resolution of hawk vision, to bat echolocation, to the twelve cones of the mantis shrimp visual system. The neurons that are not sitting in a single tangled mess of weights and nodes (akin to a single Foundation Model), but rather cleverly interconnected into neural circuits that have specific tuning to \textit{collectively} solve many tasks. Perhaps we should take inspiration from these specialized systems versus aiming for a single model that could reach artificial superhuman intelligence. Thus, we should also focus on building smart, adaptive agentic systems of specialized models.

\section*{AI Agents}

There is emerging work that I believe will make building AI agents (agentic systems) that are imbued with adaptive algorithms found in biologically intelligent systems possible. Agents that act as operating systems are one promising avenue. They can be a series of LLMs (which is currently being explored~\cite{packer2023memgpt,ye2023amadeusGPT}), but also a series of VLMs and other multi-modal language models~\cite{bordes2024introductionvisionlanguagemodeling}, which work together to ingest data, select the appropriate task-specific encoder, and link these together with downstream processing scripts. Concretely, an example of this is video behavioral analysis, where an agentic system must select the best pose estimation model for the animal type in the video, perform semantic segmentation with another model, and then perform task-programming to create ethograms~\cite{ye2023amadeusGPT}.
\medskip

To develop more adaptive agentic systems, domain-specific encoder models are essential to achieve optimal performance. As in the example above, I propose an agentic system whose architecture relies on a series of encoder modules, yet adds two new aspects (Figure~\ref{fig:agents}c). One, the agentic system that relies on a series of encoder modules, i.e., one for extracting poses, another for ingesting text plus images (VLMs), and one for extracting dynamics from neural spikes, is mediated by \textit{prediction errors} from a decoder. Two, the outputs of each specialist encoder is \textit{jointly optimized in a latent space}, which can then both learn joint representations that may be inherently more brain-like, but also a more powerful than a single Foundation Model without specialist nodes. This approach is domain-agnostic but draws on neuroscience to enhance adaptability.
\medskip 

How could this work? Encoders would be individually pretrained (think domain specific Foundation Models, such as an encoder for animal behavior~\cite{Ye2024SuperAnimal} or unified neural region-specific pretrained encoders~\cite{azabou2023a}) and their outputs are jointly optimized into another transformer (Figure~\ref{fig:agents}c). The issue becomes that while each encoder model starts as an expert, over time they need to learn more information, and the question is (1) how to sense that they need to undergo learning, and (2) how to adapt them. I propose that we could dynamically ``lock'' encoder models that show robustness on out-of-distribution samples and robustness to adversarial attacks; implying that they continually need monitored, by a dedicated LLM node, to provide trustworthy outputs. Therefore these encoders don't need to be trained (adapted) until they are deemed no longer robust. This process itself I see as similar to the cortex-basal ganglia loop that is instrumental in learning and habit formation~\cite{Graybiel2015TheSW}. Namely, these encoders could be in states of ``skill learning'' or ``habituation'' (frozen) and be constantly monitored for robustness (Figure~\ref{fig:agents}c). When robustness levels fall below an acceptable threshold, then continual learning, memory-replay, pseudo-labeling or injection of new high-quality labeled data could be added without taking the entire agentic system offline, just the encoder(s) that is not trustworthy. There has been a surge of exciting work on building OOD detection modules~\cite{gummadi2023metacognitiveapproachoutofdistributiondetection,mirzaei2025} and explainable attribution methods in time-series and images~\cite{samek2019explainable,ribeiro2016should, lundberg2017unified,Sundararajan2017AxiomaticAF,SchneiderIdentifiableAM} that could make this an attractive path forward.
\medskip

We can also use a neuroscience-inspired approach by specifically adding prediction errors into the LLM decoder that are monitoring each encoder. Effectively, prediction errors provide a signed teaching signal when an uncertain prediction is detected. The development of predictive-coding inspired networks already show promise~\cite{lotter2017deeppredictivecodingnetworks,Assran2023SelfSupervisedLF,schneider2023cebra}, and now with modern transformer architectures we could leverage prediction errors in order to do not only then do in-context learning within the encoder online, but signal which encoders need unlocked and updated -- akin to how the brain must decide to online change a motor command vs. update an internal model. Here, I take inspiration from the motor system~\cite{Hausmann2021MeasuringAM}. A one-time  error means ``in-context learn'' and change your ongoing motor command, but a repetitive error means update your internal model (Figure~\ref{fig:animals-adapt}c). If the incoming perturbation cannot be solved with online adaptation, a new internal model would then need to be created, which takes more time and can be completed ``offline'' in the unlocked model. Then methods such as continual learning, pseudo-labeling and memory-replay, or synaptic intelligence could be used to adapt the model online. Moreover, new discoveries in neuroscience should inspire new online learning techniques. The neural schema in prefrontal cortex discovered by~\citet{El-Gaby2024nature} could inspire new approaches for adaptive memory-replay and new model building as well (Figure~\ref{fig:animals-adapt}b). Models taken offline would use standard techniques such as gradient decent to fine-tune the model on the data that produced a prediction error (and therefore a drop in robustness). 
\medskip

The second aspect of my proposal is to jointly optimize a downstream joint-latent space that takes as inputs the output tokens from each specialist encoder. This could provide an embedding that has a richer representation of each sub-task vs. jointly optimizing on raw input data as in Foundation Models. Importantly, this jointly optimized space can have it's own LLM agent, with it's own prediction errors, that also provides human natural language interactions such that at inference-time the model outputs human-interpretable reasoning, aiding in transparency, trustworthiness, and interpretability. 
\medskip

Lastly, independent of the agentic system, taking the cellular diversity found in the brain seriously in neural networks should drive innovations in architecture design. SNNs have both excitatory and inhibitory, and even neuro-modulatory units compared to current transformers. Can we merge the power of large transformers and the scale of data animals receive, with the diverse cell types that produce neural computations (Figure~\ref{fig:circuits})? Perhaps gating mechanisms akin to transformer blocks (Figure~\ref{fig:agents}b) that add a signed (excitatory or inhibitory) tokenization and masking with biological time-delays can lead to new innovations in adaptive artificial systems.

\section*{Closing Remarks}

As we strive to move beyond traditional AI towards building truly adaptive intelligence (Figure~\ref{fig:agents}d), a key lies in the integration of insights from biological systems.
I argue that adaptability will emerge from incorporating neuroscience-inspired principles, particularly internal models that leverage prediction error-based updating that refines internal representations based on discrepancies between expected and observed inputs, and from the brain-like modularity that structures the system with functionally distinct yet interconnected encoders, akin to sensory and motor modules that differ in data types but share an architecture. By integrating these mechanisms, agentic systems could achieve greater robustness and generalization across tasks, making them more effective in dynamic and uncertain environments. Inherent to these ideas is that these adaptive agentic systems can also be embodied into robotic hardware to systematically test, in fully observable sensorimotor system, their robustness and generalization. Ultimately, such advances therefore not only push progress in AI, but impact neuroscience by providing a new framework and testbed for theories of the mind.
\end{multicols}

\section*{References}
\bibliography{references}

\subsection*{Acknowledgments}

I would like to thank Alexander Mathis for discussions, Shaokai Ye, Hossein Mirzaeri, Georg Keller, Markus Meister, and Travis DeWolf for providing input on an early version of the manuscript, and all my lab members who have continually shaped my interests across machine learning and neuroscience.

\subsection*{Conflict of Interest}

I declare no conflicts of interest. 

\end{document}